# Nuclear power for energy and for scientific progress

G. Giacomelli and G. Maltoni
University of Bologna and INFN Bologna
giacomelli@bo.infn.it , maltoni@bo.infn.it

**Abstract.** The Introduction in this paper underlines the present general situation for energy and the environment using the words of the US Secretary of Energy. A short presentation is made of some major nuclear power plants used to study one fundamental parameter for neutrino oscillations. The nuclear power status in some Far East Nations is summarized. The 4th generation of nuclear power stations, with emphasis on Fast Neutron Reactors, is recollected. The world consumptions of all forms of energies is recalled, fuel reserves are considered and the opportunities for a sustainable energy future is discussed. These considerations are applied to the italian situation, which is rather peculiar, also due to the many consequences of the strong Nimby effects in Italy.

## 1. Introduction

The US secretary of energy Steven Chu stated recently [1]  :
"The industrial revolution began in the mid-eighteenth century and provided humans with capabilities well beyond animal and human power. Steam-powered trains and ships, and then internal combustion engines transformed  how people moved and produced goods around the world. Electrification and related technologies continued the revolution in the nineteenth and twentieth centuries. Today a growing number of people keep their homes warm in the winter, cool in the summer and lit at night. They go to the local market in cars with power of over a hundred horses. This power is derived largely from our ability to exploit fossil sources of energy. However, in the transition from human and horse power to horsepower, the carbon emissions that result from the equivalent of over a billion horses working continuously have created significant climate-change risks." *[)]

"Carbon dioxide emissions are expected to increase from 19 gigatons per year to 43 Gt/yr." The world needs another industrial revolution in which our sources are affordable , accessible and sustainable. Energy efficiency  and conservation as well as decarbonizing our energy sources , are essential to this revolution."
"Despite the significant growth in the use of renewable energy, the fractional sum of non-carbon emitting sources of energy remained small in the past 2 decades."

-------------------------------------------------------------------------------------------------------
*[)] In the 1850s many large cities (New York, London,…) had a great problem for disposing of the waste made by the many horses used for travel purposes.

## 2. Nuclear reactors for power and for physics

" Nuclear power can have an important role in efforts to decarbonize the production of electricity. Worldwide, nuclear energy constituted about 14% of the total electrical power generated in 2009."

"Third generation nuclear-power plants are engineered to be significantly safer than previous generation ones."

" It is possible that safe nuclear power can be made accessible through the economy of constructing many reactors in a factory rather than one at a time at each site."

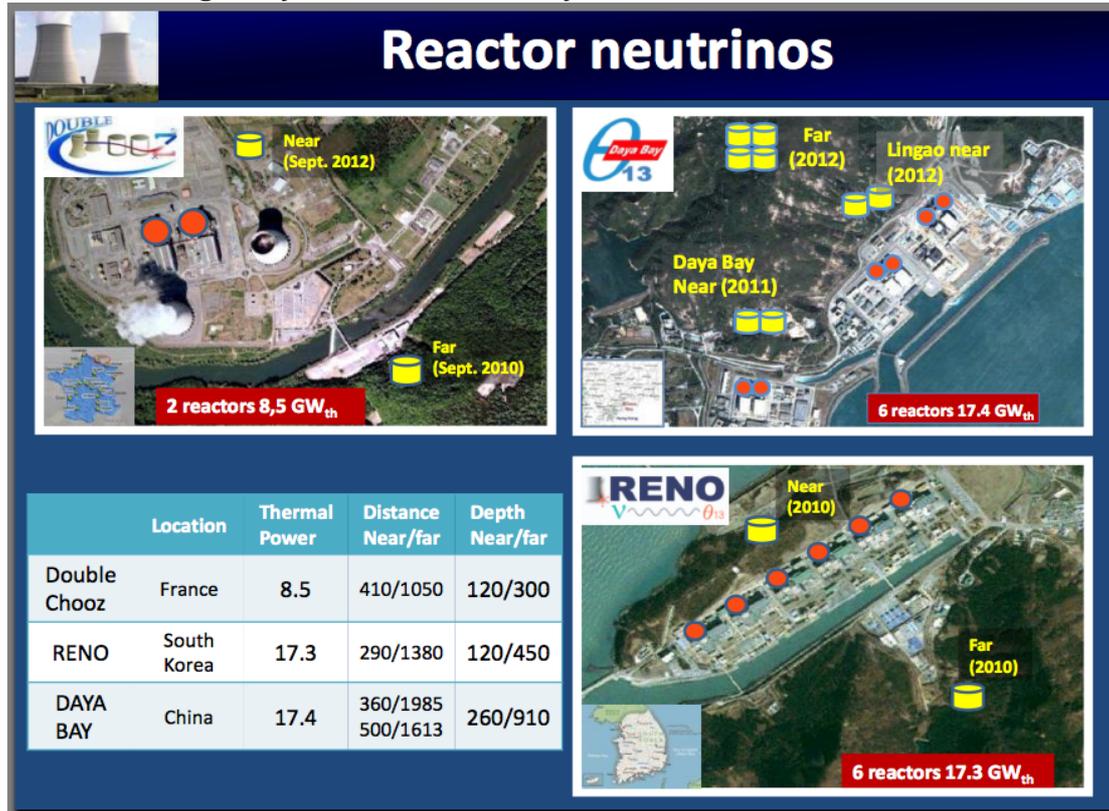

Fig. 1. The 3 nuclear reactor sites recently used for the measurement of the neutrino oscillation parameter $\theta_{13}$.

Recently 3 sites with several nuclear reactors have been used to measure the neutrino oscillation angle parameter $\Theta_{13}$, the last parameter needed for understanding the phenomenon of neutrino oscillations [2]. All the nuclear reactors are of the latest french-german design (EPR of generation 3 or $3^+$):
 -2 reactors in France (Exp. Double Chooz)
 -6 new reactors in South Korea [Exp. Reno]
 -6 new reactors in China (Exp. Daya Bay+Lingao) near Hong Kong (This is the complex of reactors with the present highest total power).

The experiments attracted large International Collaborations and obtained important results taking advantage of the very large fluxes ($10^{21}$-$10^{22}$ low-energy electron anti-neutrinos per second) from the power reactors.
$\theta_{13}$ is small but non zero : $\sin^2\theta_{13}=0.098+-0.013$ [3].

The main practical purpose of these reactors is the production of electric energy without any $CO_2$ emission. This is particularly important for China.

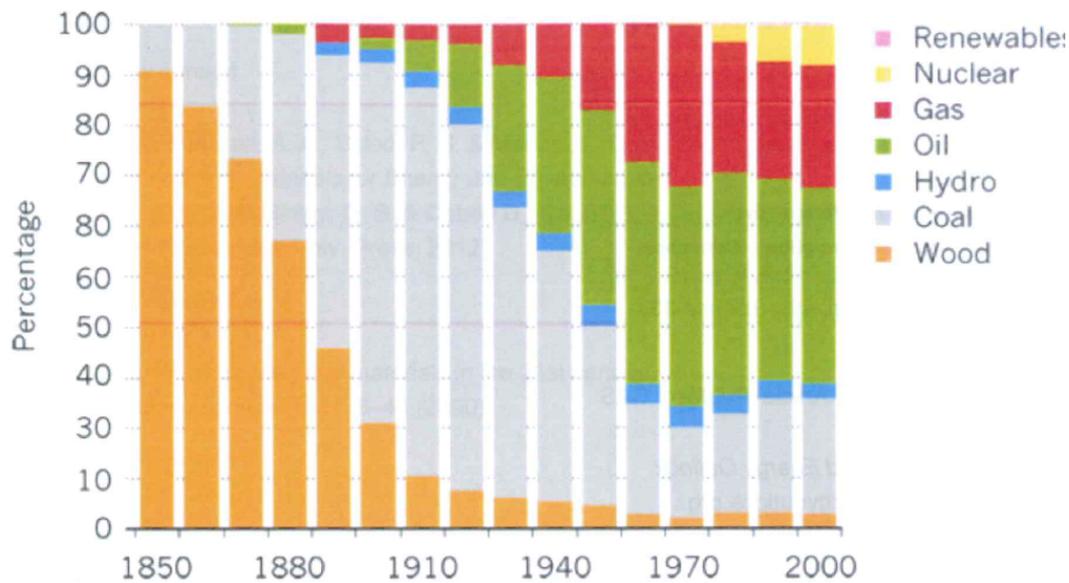

Fig. 2. Historical change of energy sources. Note that changes were made over many decades.

After the Fukushima incident in Japan, all Nations adopted safety improvements in all their nuclear reactors. While most Western Nations still hesitate to state clearly their future programs (see talks by Sumini and Toscano [4] ), most Far-East Nations decided as follows:

-Japan stopped its nuclear power stations, but recently restarted 2 reactors and plans to eventually restart its nuclear power program.

-South Korea (SK) has 23 nuclear reactors with a total power of 20.7 GWe which supplies about 1/3 of the employed electricity at a very low cost per kWh. SK plans to have, by 2030, 40 units to provide 56% of the electricity needed. It may be worth mentioning that SK won a $20 billion contract to supply 4 nuclear reactors to the United Arab Emirates (UAE).

-China: has 14 nuclear reactors of German-French and US design and 25 under construction. They plan to become self sufficient in design and in building technical parts, and to have a 5 fold increase (up to 60 GWe) by 2020, 200 GWe by 2030 and more later on.

-India is using 4 GWe nuclear power and plans to reach 14.6 GWe by 2020 (they get the enriched U fuel for their reactors from Russia).

## 3. Fourth generation nuclear reactors. Fast Neutron Reactors

The 4th generation nuclear reactors are planned to be much safer, assuming many passive automatic safety systems, in particular water cooling of the reactor core by gravity, without external electric current.

Fast neutron reactors (FNRs) are a step forward from present conventional reactors and they will yield a much more efficient use of Uranium:

-FNRs use $U^{238}$, over 99% of natural uranium

-besides $U^{238}$ , they may burn also $U^{235,}$, Pu and the actinides (most of the actual high-level long-lived nuclear wastes)

-the FNRs which produce Pu are called **breeder** reactors

-the best solution would be a closed cycle, burning everything

-at present, these reactors have an experience of 400 reactor-years and further research is now on in many Nations ( these reactors are not yet commercially ready)
-the 4th generation nuclear reactors will probably be mainly FNRs
-the world $U^{235}$ fuel supply for present nuclear reactors is estimated to last about 50 y, while the $U^{238}$ supply would last for thousands years.
-many Nations plan to use FNRs as soon as they will become available.
-France stated that half of their present reactors will eventually be replaced by FNRs
-China has one 20 MWe experimental prototype
-India has one 500 MWe experimental prototype
-Other prototypes are in USA, UK, Germany, Japan, Kazakhstan, Russia.

Much research is being performed on **Thorium Reactors**, expecially in India. Thorium is more abundant than U, and India has much Th. An advantage of Th is that no nuclear bomb can be made with it. But not all the problems are solved yet.

## 4. Nuclear Energy in Italy

In Italy, before the 1964 nationalization of nuclear energy, three nuclear reactors were working, or in an advanced construction stage: in Latina (200 MWe) built and controlled by SIMEA (AGIP and IRI), which started operations in 1963: the PWR Garigliano reactor by IRI ((164 MWe), operative since 1964, and Trino Vercellese (TRINO1), a 160 MWe PWR, by Edison. After nationalization, they passed to ENEL, that made operative in 1984 the 4th nuclear reactor (the last one) at Caorso, a BWR of 860 MWe. After the 1987 referendum it was decided to stop the construction of more reactors, the immediate stop of Latina, Caorso and Trino 1, and to convert the 2 almost completed reactors at Montalto from nuclear to methane (but the italian sea ports in the Tirrenian sea refused to accept the Algerian liquefied gas, which was then sent to Marsiglia, from where it arrived to Montalto in a gas pipeline) .

The cost for quitting nuclear energy was estimated by ENEL at approximately 120.000 billions of the old lire. Adding the increased costs of oil and gas, the total cost mounted to about 100 billion euros and now it must be added the planned cost of 9.2 billion euros for decommissioning all the italian sites, as stated below [4].

In order to optimize the cycle and to reduce the nuclear waste, the preference is now for adopting the close cycle for all reactors, recuperating $U^{238}$, unfissioned $U^{235}$, Pu and actinides, which are recycled to make new nuclear combustion material (MOX).

The nuclear waste of normal $U^{235}$ reactors is always rather small in size compared to any other type of waste, and it becomes almost negligible after recicling with MOX. It would become completely negligible using FNRs.

-**Dismantling of italian nuclear reactor sites:** "La piu' grande bonifica nella storia del Paese"[4]

"The italian company Sogin spent **1.7** billion euros in the last 10 years to start the dismantling of the italian nuclear sites, and terminated, until now, the dismantling of Caorso turbine for which they made 77000 radiological checkings under Ispra and Arpa Emilia-Romagna control ( hoping that the radiological measurements were more efficient and less burocratic than those made at Montecuccolino, Bologna, where for a few sub-critical units turned off few years before, the inside radioactivity became lower than the outside natural radioactivity). Sogin plans to complete the decommissioning of all the italian sites by **2026** adding **5** more billion euros. Moreover they plan to make a technological parc and the national deposit for low intensity nuclear wastes produced by hospitals and medical Institutions, at an extra cost of **2.5** billion euros (total cost **9.2** billion euros) [4]".

It seems a well planned but very costly operation, as if it were for the construction of a new plant, with lots of radiological measurements, emphasis on long range employment and no apparent money saving consideration; also the title seems very ambitious. The total cost seems to correspond to nuclear reactors operative for 40 y. In other nations, like France, these expenses are recovered from the money set aside for each operation year of the nuclear plants. In Italy all reactors were operative for much shorter times. Most of the anticipated cost seems to be included in the ENEL electric bills paid by all individual users and all industries. The italian electric bill is one of the highest in Europe. Aluminium industries, like the Alcoa in Sardinia, need large quantities of electric energy and would clearly benefit from a cheaper electricity price: the inclusion of the foreseen high dismantling costs of nuclear power plants in the electric bills is dangerous for these types of industries.

About one half of the Sogin dimantling cost could have been sufficient for making at least one site ready for a $4^{th}$ generation nuclear plant, built by italian companies or in cooperation with reliable European or Far East Nations.

Italy imports 12.5% of its electricity from the equivalent of 6 french nuclear 1000 MWe fission reactors and will pay the European Union tax for $CO_2$ emissions, while France and other nations will benefit for producing electricity (even if sold abroad) without emitting $CO_2$ [5][6][7].

## 5. World Energy Consumption

All the energy available on Earth comes from: i) solar energy, ii) natural radioactivity (nuclear energy), geothermal energy, and energy from tides. The vegetables on Earth convert the arriving solar light, via photosynthesis, into biomass with an efficiency of about 1%. The solar power used by the plants in their photosynthesis is about 1 TeraWatt, about 6 times more than the amount used by men and women now. This process, used for geological times, produced the large quantities of fossil biomass as oil, coal and gas.

Fig. 3 shows the world consumption from all available sources, which are mainly fossil fuels. Considerable power comes from hydroelectricity and from nuclear reactors. Notice that the contribution from renewable energies is growing, but is still rather small.

Fig. 4 shows the distribution of proven oil reserves in 1991, 2001 and 2011 [2]: notice that we do not have any peak versus time, as was hypothesized some time ago. The reserves seem to be increasing with time : this is due to continuosly improved technologies. The actual reserves should secure oil use for more than 50 years. In the future oil and gas from oil sands and probably from oil shales will be used, if prices will not be too high (see Fig. 5) [2, 5, 8, 9, 10, 11].

The BP statistics show that the gas reserves plotted versus time indicate a tendency very similar to the oil reserves of Fig 4. Also this is interpreted as due to the improving technologies.

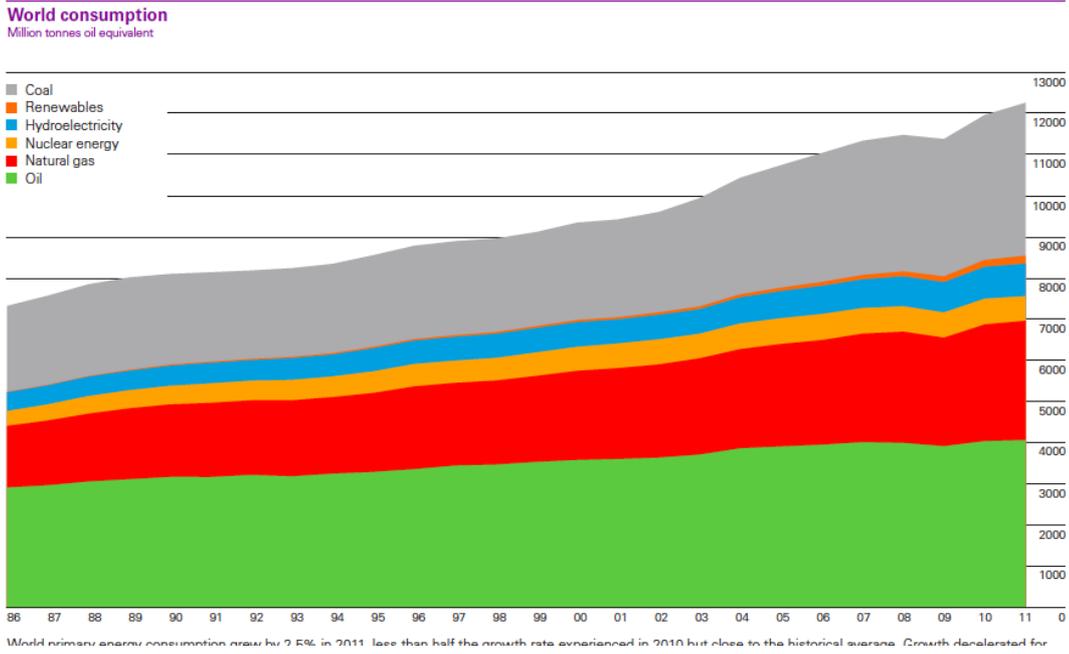

Fig. 3. World energy consumption from all available sources, which are mainly from fossil fuels. Some power comes from hydroelectricity and from nuclear reactors. The contribution from renewable energies is still rather small [2].

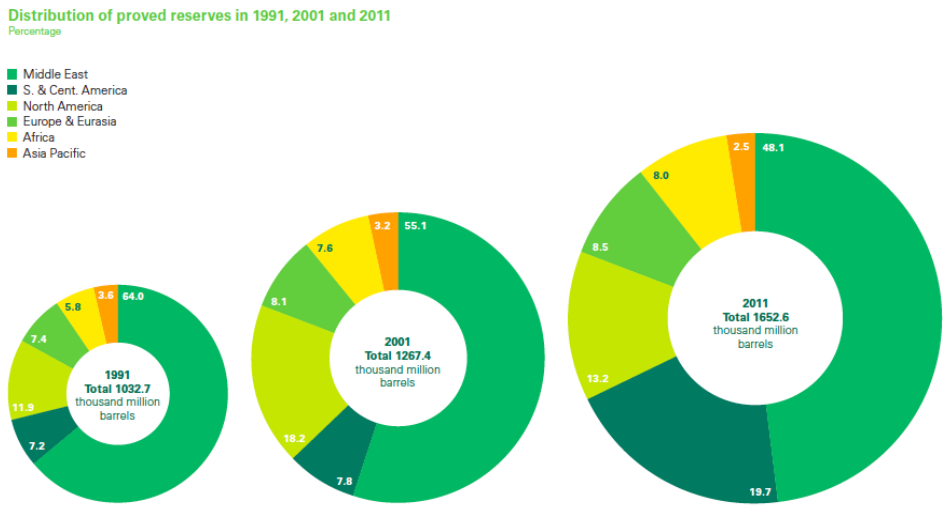

Fig. 4. Distribution of proven oil reserves [2]. In the future one may use oil and gas from oil sands and from oil shales.

   The BP statistics for coal reserves show a different trend, which is primarily due to a new classification of the types of coal : there is no effective problem and the world still has large quantities of coal [2].

# Fuel reserves (billion tons of oil equivalent) ▶Ravenna 2012

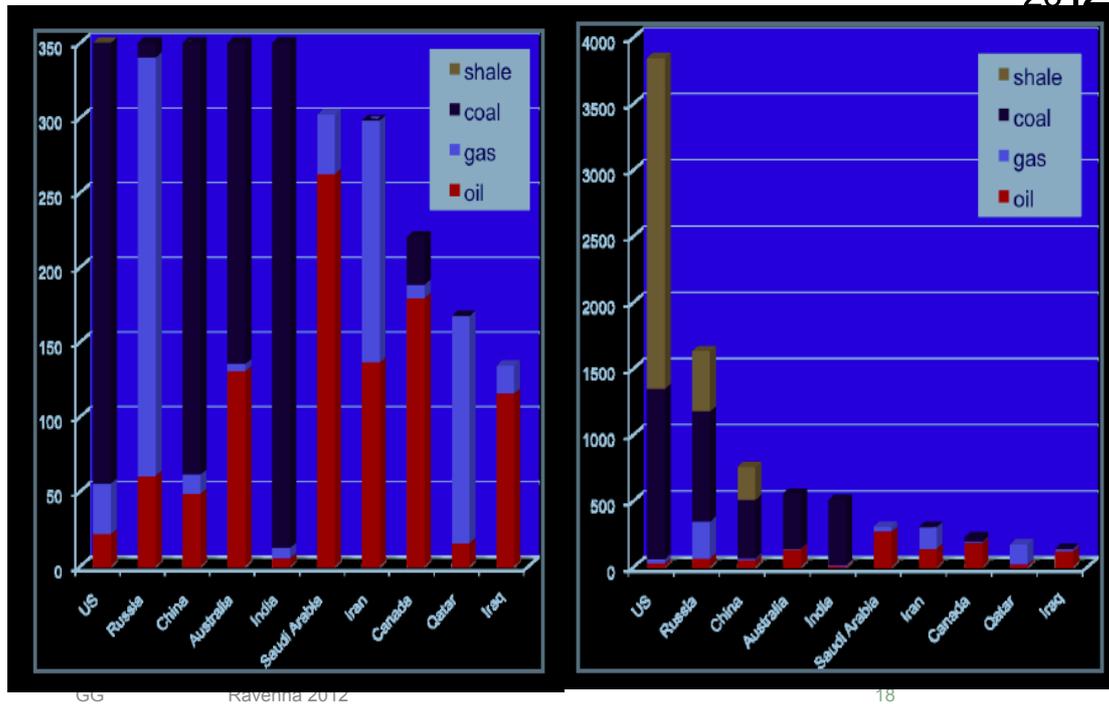

Fig. 5. a). The oil+gas reserves assuming the use of oil sands, as is partly done in Canada (which would have oil reserves almost as large as Saudi Arabia). b) The oil+gas reserves assuming the use of oil shales (at larger costs) : the US could become the country with the largest reserves of oil and gas !

## 6. Renewable Energies

Fig 6 shows a pictorial summary of almost all renewable energies.
We should remember that "Our ability to find and extract fossil fuels continues to improve. Economically recoverable reservoirs are likely to keep up with the rising demand for decades. The stone age did not end because we ran out of stones (We only transitioned to better solutions [1])."
The cost of renewable energy is becoming competitive with other sources of energy. Western Europe and the USA tend to accelerate the transition to affordable and sustainable energies that will improve economic growth, increase energy security and mitigate the risks of climate changes. They suggest to improve and use any new type of renewable energy and perform all possible energy savings. The suggestions may be somewhat optimistic. On the other hand, if we do not change direction soon, we may end up where we are heading.
   -The IEA says that wind energy will be highly competive by year 2020: wind costs decline with new blades, increased sizes, and higher heights.
   -Also the cost for photovoltaic is dropping below the "power law experience curves", and solar concentrators should improve the performance and further reduce costs. Thin photovoltaic panels of relatively high efficiency are becoming available. But for the moment most solar panels are made in China.
   -The yield of sugar cane in Brazil increased by a factor of 3 in 30 years, see Fig. 7b.

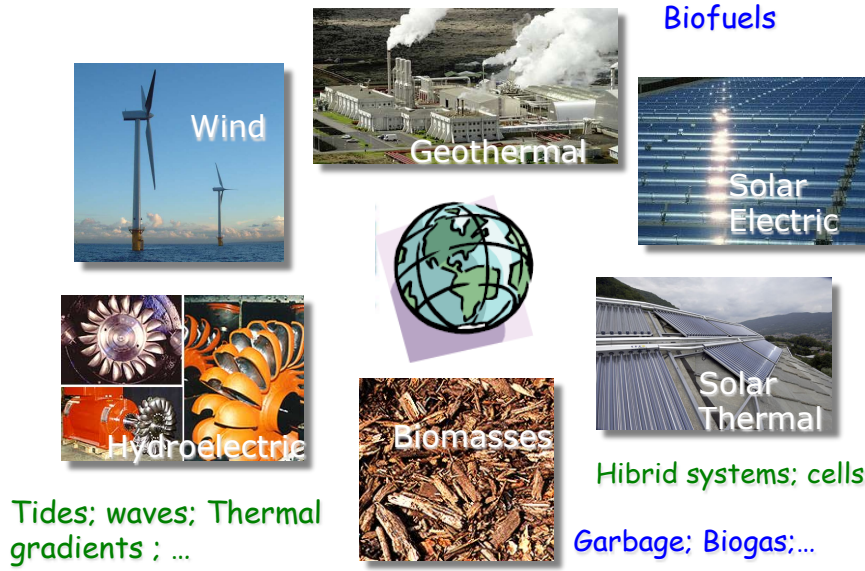

Fig. 6 : Illustration of Renewable Energies.

 -New ways of financing could help to reduce the problems encountered in the development of renewable energies.
 -A small improvement in Italy : The alga lactuca was a strong inquinant in the northern Adriatic Sea and specifically in the Venice laguna. The firm Cartiera Favini in Rossano Veneto managed to collect and reprocess the algae to produce paper, ecological paper. The yield of algae is much larger than the yield of vegetables produced on land. The use of algae for making paper saves a large number of trees. Now the Cartiera Favini is capable to reprocess different types of agronomic materials (apples, oranges, etc) for the production of ecological paper substituting cellulose from trees [12]. One hopes that this leads to some energy savings.

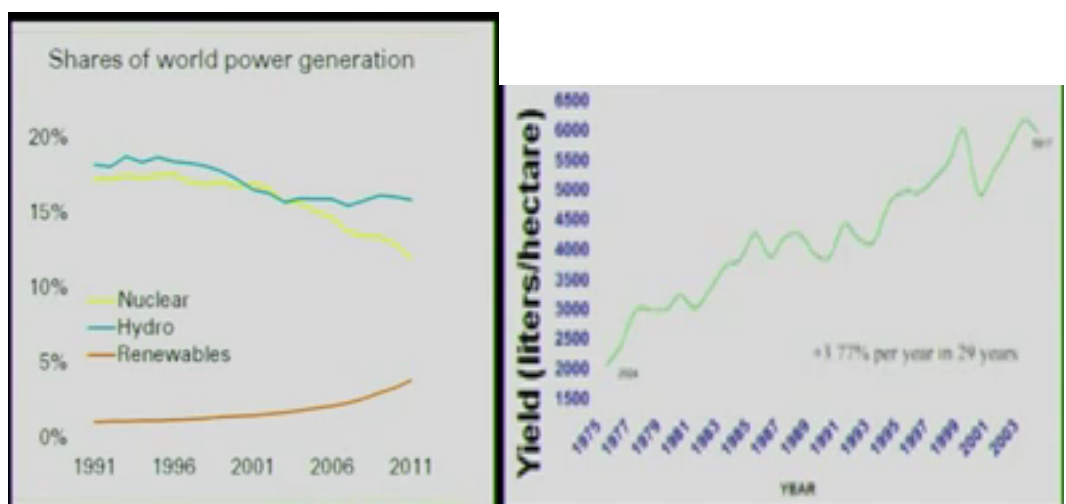

Fig. 7a) Renewable energies grew considerably in 1911-2011, but their global impact is still small (red line); Nuclear and Hydro power declined and the fraction of carbon-free primary energy remained about constant. b) In Brasil sugar cane yield of ethanol per hectare increased by a factor of 3 in 30 y.

## 7. Energy savings

"The term *crisis* in chinese is made of two words : the first means *danger*, the second means *opportunity*: one may thus be encouraged to take advantage of the many opportunities which may arise in a crisis [1]."

Very many energy saving programs have been studied and are slowly applied.

In the short term Energy Savings may be considered as a new source of energy.

We shall quote a few types of energy savings:

-Electric energy savings with the use of low consumption lamps: it is already being implemented on a large scale.
-Large energy savings were obtained with the new airplanes: they are lighter, have more efficient engines (50% since 1970). Further improvement are obtained with the largest planes like the new Airbus 380.
-The use of lighter materials for cars (carbon fibers and titanium) and/or better aerodynamics ( including better aerodynamics for the bottom parts of trucks) may soon allow an effective reduction of fuel consumption (5-10%). A reduction of the friction is possible, may be 20% in the short term, and even 60% in 15 years. An additional improvement in gasoline engines (25% ?) should be possible with the use of higher octane fuels.
- Carbon micro/nano fibers injected and polimerized later: together with new simulations should reduce the costs and reduce car weights.
-The 3rd generation steel production may yield new steel with a higher strength.
-New metal alloys together with computer simulations may allow an increased use of lighter materials.
-Electric vehicles and hybrids could further improve the traffic situation.
-The energy densities of competing batteries: research may improve the efficiency.
-Electric Power conversion : most future power should be transported through power electronics. which may allow a considerable energy saving.
-There may be a lowering in the fuel chemical precursors and in the cost of biomass harvesting, storage and shipment in Brasil and in the US. But most Nations should be careful that these methods do not increase the cost of food nor produce a shortage.

## 8. Conclusions

Fig. 8 shows the new Colorado river bridge near the Hoover Dam on U.S. route 93 between the states of Arizona and Nevada and the cities of Phoenix and Las Vegas. The Hoover Dam was built during the recession of the 1930's, and it became a tourist attraction with about one million visitors per year. The new Colorado river bridge, at a considerable height, was constructed starting at the beginning of the recent financial crisis and, using new techniques, remained within the approved budget, was operating on time in 2011, and started to be another effective tourist attraction.

These successes may be considered as examples of beating the NIMBY effect (Not In My BackYard), which is causing so many delays and extra costs in many major projects in Italy ( in several hundred projects). Some examples are the high velocity connection (TAV) between France and Italy in Val di Susa, the power station of Porto Tolle, the regassifier of Brindisi, many industrial projects, etc. The NIMBY effect involved also the networks of radio emitters used by portable phones. But in this case the italian love for portable phones seems to have cancelled this effect!

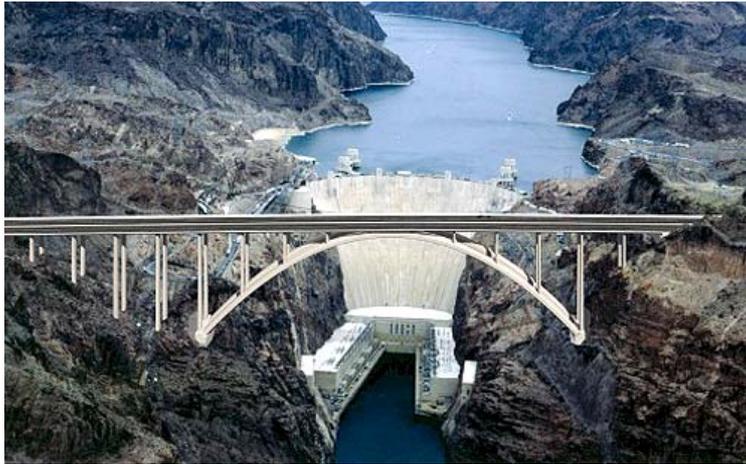

Fig. 8. The new Colorado river bridge on U.S. route 93 near the Hoover Dam.

In any case we may need a cultural change if we do not want to be left technologically behind by the emerging nations like China, India, Brasil, South Korea and many others.

**Acknowledgements.** We would like to thank the participants to the Ravenna 2012 workshop on "Fare i conti con l'ambiente. Rifiuti, acqua, energia", in particular we thank the organizers headed by prof. L. Bruzzi.